\documentclass[preprint,12pt]{elsarticle}
\usepackage{graphicx} 
\usepackage{textcomp,mathcomp}
\usepackage{siunitx}
\usepackage{float}
\usepackage{subcaption}
\usepackage{svg}
\usepackage{color,soul}
\usepackage{hyperref}
\usepackage{svg} 
\usepackage{placeins}

\usepackage{xcolor}	
\usepackage[normalem]{ulem}
\usepackage{soul}															

\begin{document}


\begin{frontmatter}
\title{Sensitivity increase of 3D printed, self-sensing, carbon fibers structures with conductive filament matrix due to flexural loading}

\affiliation[inst1]{organization={University of Twente, Robotics and Mechatronics group}, addressline={Drienerlolaan 5}, postcode={7522 NB},  city={Enschede}, country={the Netherlands}}

\affiliation[inst2]{organization={University of Twente, Production Technology group}, addressline={Drienerlolaan 5}, postcode={7522 NB}, city={Enschede}, country={the Netherlands}}

\author[inst1]{Matei Drilea}
\author[inst1]{Alexander Dijkshoorn}
\author[inst2]{Gusthavo Ribeiro Salom\~{a}o}
\author[inst1]{Stefano Stramigioli}
\author[inst1]{Gijs Krijnen\corref{cor1}}
\ead{gijs.krijnen@utwente.nl}
\cortext[cor1]{Corresponding Author:}

\begin{abstract}
The excellent structural and piezoresistive properties of continuous carbon fiber make it suitable for both structural and sensing applications. This work studies the use of 3D printed, continuous carbon fiber reinforced beams as self-sensing structures. It is demonstrated how the sensitivity of these carbon fiber strain gauges can be increased irreversibly by means of a pretreatment by pre-stressing the sensors with a large compressive bending load. The increase in the gauge factor is attributed to local progressive fiber failure, due to the combination of the thermal residual stress from the printing process and external loading. The coextrusion of conductive filament around the carbon fibers is demonstrated as a means of improving the reliability, noise and electrical connection of the sensors. A micrograph of the sensor cross section shows that the conductive filament contacts the various carbon fiber bundles. All-in-all, the use of pre-stressing carbon fiber strain gauges in combination with coextrusion of conductive filament hold promises for 3D printed structural sensors with a high sensitivity. 
\end{abstract}




\end{frontmatter}

\section{Introduction}{\label{intro}}

Additive manufacturing technology has achieved widespread use in research and manufacturing, enabling the production of complex parts which oftentimes cannot be made with traditional manufacturing techniques~\cite{AM_intro}.  One aspect holding back this technology is related to the usable materials which often do not posses the strength to be suitable for engineering grade parts~\cite{AM_weak}. To solve this issue, composite materials can be used to strengthen 3D printed parts~\cite{CFR_usage, newer_review,CHENG2023}, which already has seen application in aviation, automotive, and robotics parts~\cite{CF_aero, CF_automotive, CF_robotics}. One such technology for reinforcing 3D printed structures is by continuous fiber coextrusion~\cite{Composer_Manual, Anisoprint_paper}, which is a form of Material Extrusion (MEX) that combines pre-impregnated continuous carbon fibers with a thermoplastic, which is extruded layer-by-layer to create the final structure.

Carbon fiber is an excellent material for reinforcing 3D printed structures, due to its high stiffness, high tensile strength and high strength to weight and stiffness to weight ratios~\cite{newer_review,CHENG2023}. Additionally, due to the piezoresistive properties of carbon fiber bundles~\cite{CF_piezoresistive,Chung} they can serve a dual purpose: reinforcement of the printed part and providing a means for embedded strain sensing, enabling self-sensing structures. Research on the mechanical properties of continuous 3D printed carbon fiber structures has showcased the improved strength~\cite{CF_flexural,CFR_printing_weaken}. The sensing properties of continuous carbon fiber have already been demonstrated by several works~\cite{Jonathan, Luan, new_piezo, Yao, Billah}, with applications ranging from failure detection to strain and load measurement. Heitkamp et al.~\cite{new_piezo}, Yao et al.~\cite{Yao} and Billah et al.~\cite{Billah} all 3D printed specimen with continuous carbon fiber reinforcement and measured the resistance while slowly increasing the flexural or tensile mechanical load until failure occurred. In all cases a high increase in the gauge factor of the samples was observed in the plastic deformation region before total failure. In this phase progressive damage occurs, e.g. failure of single fibers or bundles, until total failure~\cite{new_piezo}. An irreversible resistance change indicates (micro-) damage of the fibers~\cite{Chung}, as was proven through computed tomography analysis~\cite{Chang2025-he}. Furthermore, the sensing behaviour due to fracture has a signficant strain rate dependance~\cite{LIU2025}. However these studies did not utilise the irreversible changes after high loads for sensing applications. 

These examples show that there is room for new avenues of research, such as permanently changing the gauge factor by applying high loads, potentially creating highly sensitive sensors. Alternatively, the use of carbon-based conductive filaments for 3D-printed, piezoresistive sensors has been demonstrated before \cite{Banks_review,Schouten2020}. These sensors don't have the favourable mechanical properties and relatively low resistivity of continuous carbon fibers, but they can operate robustly for significantly larger strains. In case the increase in gauge factor is governed by cracks in the fibers, a conductive filament can be used as matrix material to allow for a highly resistive current path in parallel to increase the reliability and robustness. A parallel resistive pathway  will maintain electrical contact between cracked fiber segments, ensuring a relatively high sensitivity and wide sensing range, similar to crack-based sensors~\cite{zhou2022}.   
This research therefore aims to create highly sensitive continuous carbon fiber strain sensors by methodologically introducing fiber fracture. and to investigate the mechanisms through which their sensitivity is altered. The coextrusion of conductive filament with carbon fibers will also be studied in an attempt to increase the reliability and robustness of the sensors. 3D printed samples will be loaded with increasingly high flexural forces to investigate the (changes in) gauge factor when operated at lower loads. This process of loading the samples with high forces with an intent to increase their sensitivity, will be referred to in this paper as pre-stressing. In section~\ref{method}, the used materials and methods will be shown, detailing how the samples will be pre-stressed through flexural bending. In section~\ref{theory}, the theoretical basis of the paper is presented. Classical beam theory including residual thermal stress and electrical analysis will be used to model the samples. 
The results in section~\ref{results} show that the the samples can be pre-stressed, irreversibly increasing their sensitivity and resistance.


\section{Materials and Method}{\label{method}}

\subsection{Sample Fabrication}

\begin{figure}[h]
    \centering
    \includegraphics[width=0.9\textwidth]{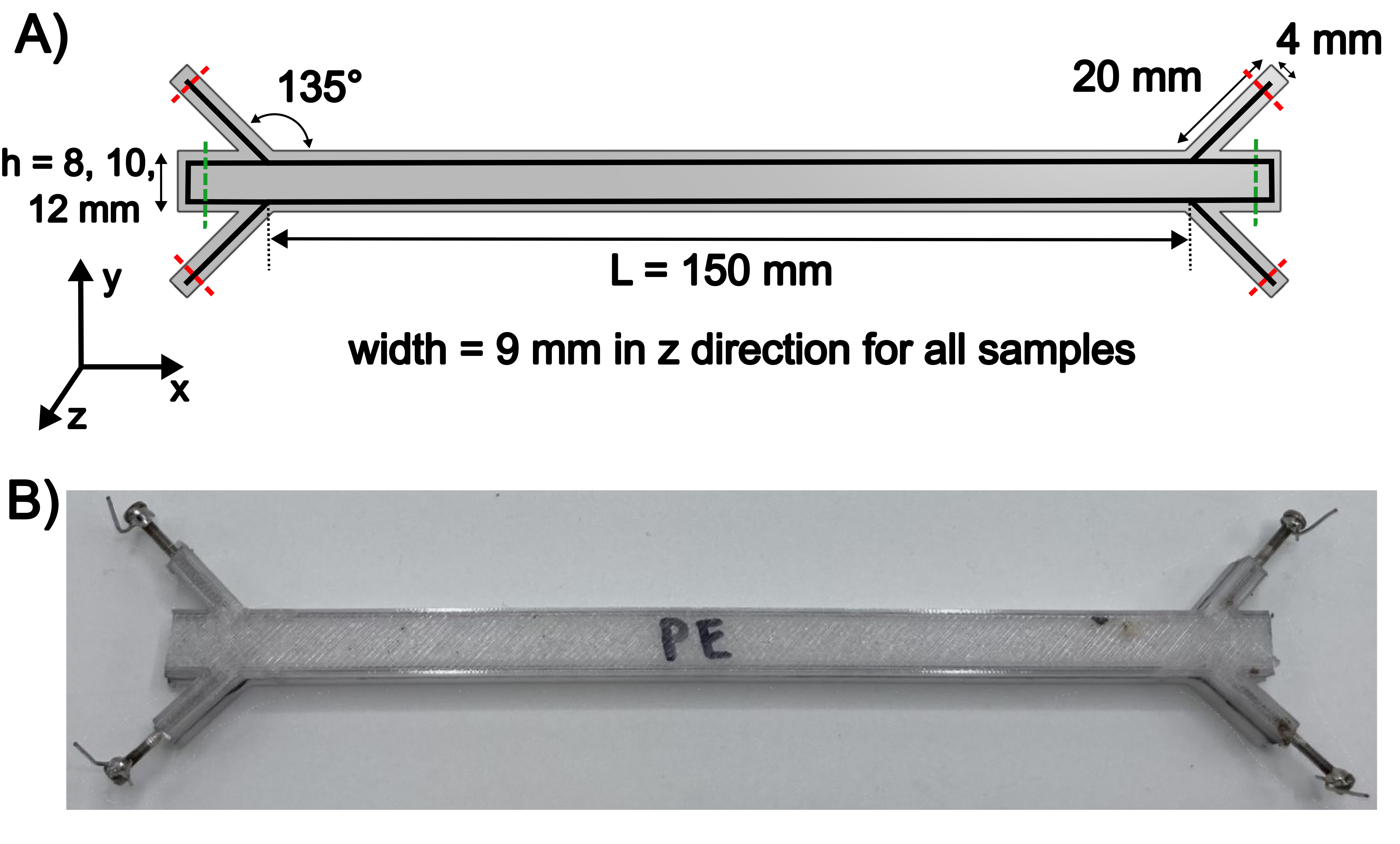}
    \caption{A. Sample design with dimensions, where the thick black lines represent the carbon fiber and the green and red lines indicate sawn off ends. B. 3D-printed sample with PETG, with sawed off ends and electrical connections.}
    \label{fig:samples}
\end{figure}

To fabricate the samples, the Anisoprint A4 Composer 3D-printer was used~\cite{Composer_Manual}, which is capable of continuous carbon fiber reinforced MEX. It achieves this by having two extruders: a regular MEX extruder and a composite extruder, which coextrudes the pre-impregnated continuous carbon fibers with a thermoplastic. The used materials are the Anisoprint CCF (Composite Carbon Fiber) 1.5K~\cite{CCF_Datasheet} and PETG filament (Polyethylene terephthalate glycol). Anisoprint's Aura 2 is used to slice the model~\cite{Aura}, where table~\ref{tab:printing_parameters} presents a list of the used printing parameters.

\begin{table}[H]
\caption{Important printing parameters }
	\centering
     \footnotesize
	\newcolumntype{?}[1]{!{\vrule width #1}}
    \renewcommand{\arraystretch}{1.2} 
	\begin{tabular}{c c}
		\hline
		\textbf{Printing Parameter} & \textbf{Value} \\
		\hline
        PETG Nozzle Temperature & \SI{240}{\celsius}\\
        Protopasta Nozzle Temperature & \SI{225}{\celsius}\\
        Heated Bed Temperature & \SI{60}{\celsius}\\       
		Macrolayer height / Coextruded fiber composite layer height & \SI{0.3}{\milli\meter}\\
            Microlayer height / PETG layer height & \SI{0.15}{\milli\meter}\\
		Flow multiplier & 1\\		
		Extrusion width of PETG & \SI{0.65}{\milli\meter}\\
        Extrusion width of coextruded fiber composite & \SI{0.78}{\milli\meter}\\		
		Outer reinforced perimeter count & 1\\		
		Inner reinforced perimeter count & 0\\		
		Plastic perimeters outside fiber & 1\\		
		Plastic perimeters inside fibers & 2\\		
		Top/bottom solid layer count & 3\\		
		Infill Density & \SI{0}{\percent}\\	
		Fiber extrusion speed & \SI{6}{\mm/\second}\\		
		Plastic extrusion speed & \SI{50}{\mm/\s}\\
		\hline
	\end{tabular}
	\label{tab:printing_parameters}
\end{table}

In order to place straight CCF bundles along the print, carbon reinforced perimeters are used at certain heights. Two single-perimeter carbon fiber composite layers are placed in the middle (between $z=\SI{4.2}{\mm}$ and $z=\SI{4.8}{\mm}$) of the sample, which is printed on its side. During loading, the sample is rotated \SI{90}{\degree} around its $x$-axis, and the carbon fiber reinforced walls become the top and bottom of the sample, forming a sandwich structure. This provides high stiffness and strength in the plane of bending, and places the strain gauges far away from the neutral line. 

The printing process results in some undesirable features, such as inconsistent fiber placement at the start and end of the fiber paths, as well as fiber placement in loops (a consequence of the reinforced perimeter setting of the slicer, connecting the strain gauges). For this reason, and to establish the electrical connection with the printed carbon fibers, additional work needs to be done on the samples post printing. 

The printed sample is sawn off in a total of six places, as indicated by the dotted lines in figure~\ref{fig:samples}a. The red lines indicate cuts that expose the carbon fibers, so an electrical connection can be established. The cuts are necessary to reach the inside carbon fibers due to the chemical stability of PETG~\cite{PETG_chemical_stability}, which makes chemical etching unfeasible. Meanwhile, the green cuts are made in order to split the carbon fiber loop into two isolated strain gauges. The cuts also solve the problem of inconsistent fiber placement, as those regions are removed.  

The electrical contacts are established by mechanical exposure and the use of conductive paint, as described by Scholle and Sinapius~\cite{CF_piezoresistive}. The sawed-off surfaces at the red lines are lightly sanded in order to increase the fiber exposure. Screws are tightly screwed into the holes, providing a solid anchor point for wires to be soldered onto. Silver paint  (Electrolube SCP26G) is placed on the sanded surface and the base of the screws, establishing an electrical connection between the carbon fibers and the screws. The screws act as strain relief for the silver paint.
A final sample can be seen in figure~\ref{fig:samples}b, with each sample having two parallel but separate strain gauges running along the length of the sample. 

In total, 6 samples were printed: three different sample heights ($6, 8, \SI{10}{\mm}$) with two fiber co-extrusion materials; PETG and Protopasta~\cite{Protopasta} (a conductive Polylactic Acid (PLA) filament). The sample names are based on the material of co-extrusion: PETG and PR (short for Protopasta), and their height: Short (\SI{6}{\mm}), Medium (\SI{8}{\mm}), and Tall (\SI{10}{\mm}). For example, PR-Short is a \SI{6}{\mm} sample co-extruded with Protopasta. In order to keep track of the orientations of the samples, a marking is made on the samples.

\subsection{Experimental Setup and Procedure}
\begin{figure}[h]
    \centering
    \includegraphics[width=0.95\textwidth]{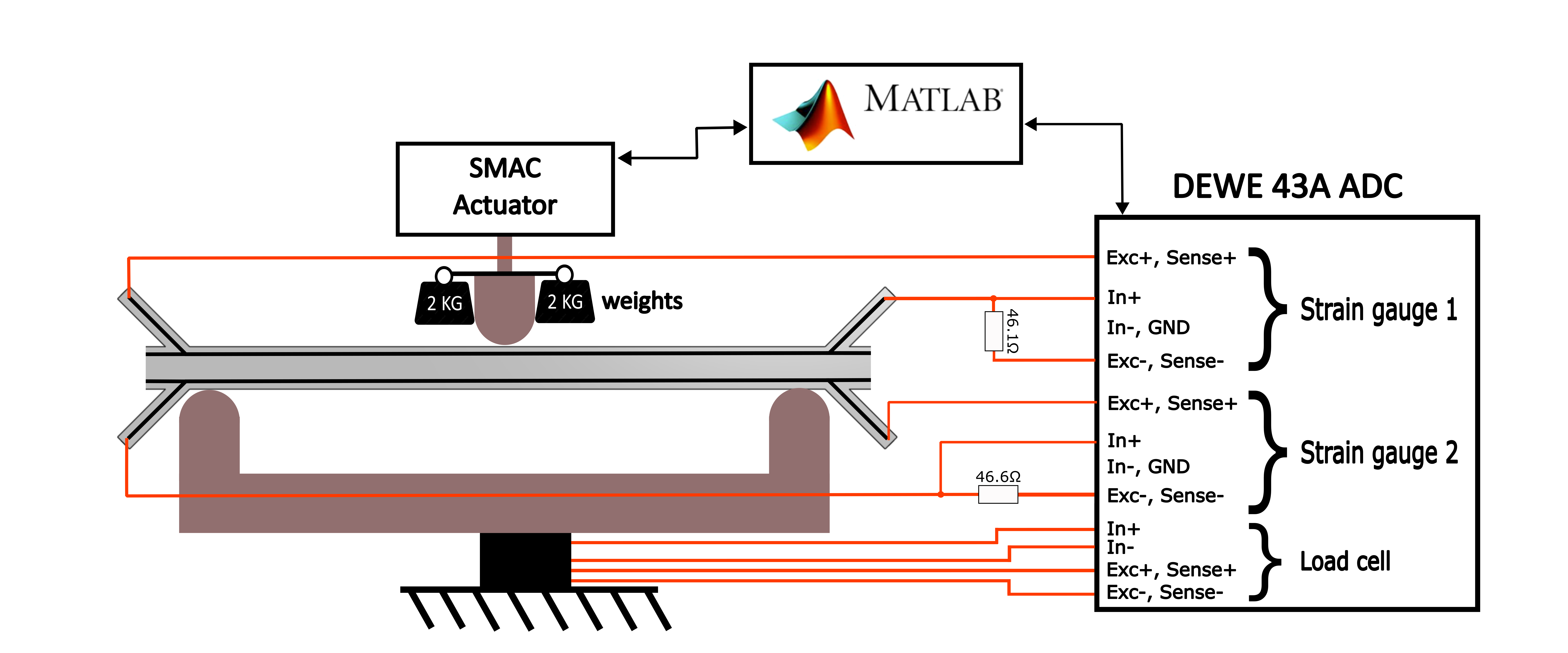}
    \caption{Experimental setup to characterize the electromechanical sensor response. For the Medium and Tall samples, \SI{4}{\kg} of additional weight is attached to the linear actuator. For the Short samples, \SI{3}{\kg} of additional weight is used.}
    \label{fig:setup}
\end{figure}

In order to characterize the electrical response of the samples under strain, a three-point bending test is used. With the samples supported on both ends as shown in figure~\ref{fig:setup}, a  SMAC linear actuator (LDL 40-100-31-3F) is used to push down on the center of the samples. To increase the maximum force of the actuator, weights are attached to the end effector of the linear actuator, increasing the maximum force exerted on the samples to \SI{70}{\N} for the Medium and Tall samples, and \SI{62}{\N} for the Short sample. A load cell measures the force exerted on the sample, and the center deflection of the beam is measured directly with the SMAC linear actuator. The deformation of the load cell itself was calibrated for and removed from the measurements. To measure the resistance of the strain gauges, a DEWE 43A Data Acquisition System is used. Each strain gauge is measured on a separate DEWE channel, using a voltage divider configuration with \SI{46.1}{\ohm} and \SI{46.6}{\ohm} resistors for each strain gauge, respectively. This allows simultaneous measurement of the resistances of the two strain gauges and the load cell, as seen in figure \ref{fig:setup}.

\begin{figure}[h]
    \hspace{-1.5cm}
    \includegraphics[width=1.2\textwidth]{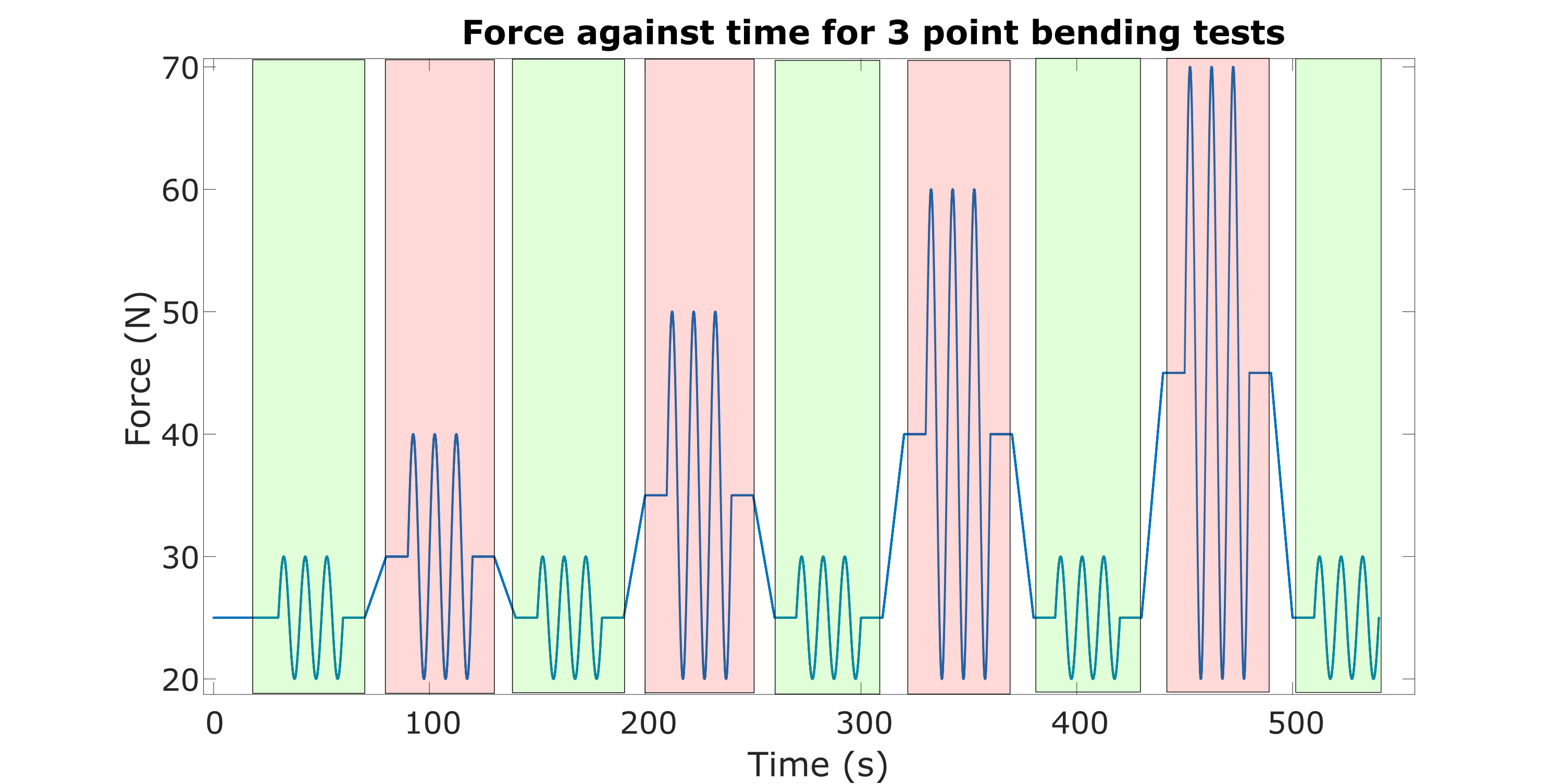}
    \caption{Visualization of the force waveform applied on the samples during the three-point bending test. The green regions measure the sensitivity of the sample, while the red regions pre-stress the samples.}
    \label{fig:force_wave}
\end{figure}

The samples were loaded in a systematic way with a force waveform. Cycles of small and large loads are applied in succession to study the effect of large loads on the sensitivity. An illustrative waveform is shown in figure~\ref{fig:force_wave} for visualization purposes. The waveform has several components:

\begin{enumerate}
    \item A set of twenty \SI{10}{\N} sine waves for measuring the gauge factor of the sample. The strain caused by these small waves is under the yield strain even for the smallest samples. The gauge factor is measured from a linear fit of the resistance against strain relationship of the entire set of small amplitude waves. The gauge factor is only measured in the small strain ranges as suggested by Chung~\cite{Chung}, and not during the set of high amplitude sine waves.
    \item A set of twenty offset, higher amplitude sine waves for pre-stressing, whose goal is to load the beam sufficiently in order to cause an irreversible increase in gauge factor. Both the offset and the amplitude of the sine waves increase progressively throughout the experiment, allowing progressive failure of the fibers.
    \item Constant amplitude waiting periods of \SI{60}{\second} to allow for some relaxation of viscoelastic effects.
\end{enumerate}

The samples are placed on the testing setup and undergo the loading as described above, referred to as the initial orientation. After undergoing the complete force waveform, the samples are flipped and loaded again with the same waveform, referred to as the flipped orientation. In the initial orientation, strain gauge 1 experiences compression and strain gauge 2 experiences tension, as seen in figure~\ref{fig:setup} and vice versa in the flipped orientation.
\section{Theory}{\label{theory}}

\subsection{Mechanical Model}
\begin{figure}[h]
    \centering
    \includegraphics[width=0.65\textwidth]{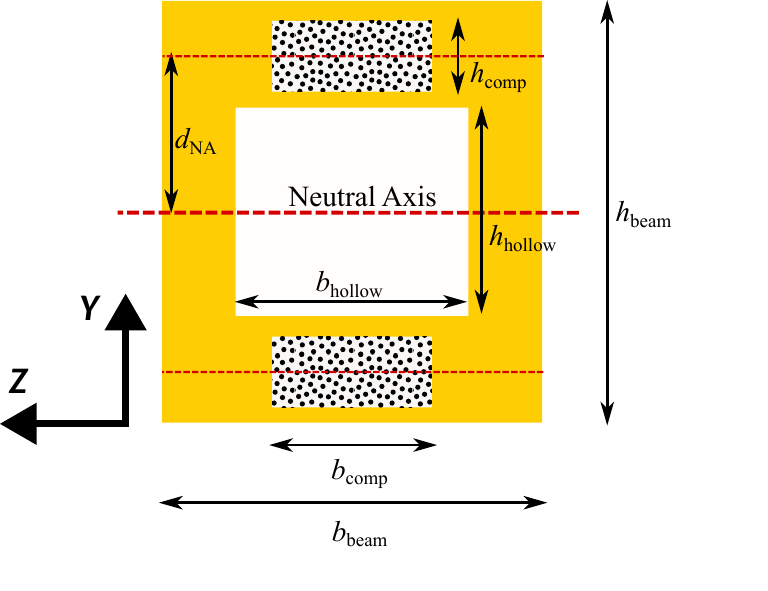}   
    \caption{Cross-section of beam, showing the carbon fiber monofilaments in white, encased by the Anisoprint resin and the co-extruded Protopasta/PETG depending on the sample (black). The rest of the structure is made up of PETG extruded from the plastic nozzle (in yellow) and a hollow core (in white). }
    \label{fig:beam_cross}
\end{figure}

The printed samples are modelled using classical beam theory~\cite{Hibbeler_MoM}. The cross-section of the printed beams can be seen in figure~\ref{fig:beam_cross}. The extruded lines of carbon fiber are made up of a combination of the pre-impregnated carbon fiber bundle (with a bisphenol-A-based epoxy resin encasing hundreds of thin monofilaments) from Anisoprint and the Protopasta (PLA with carbon black for conductivity) or PETG matrix depending on the samples. They will be modelled together as a composite material, with the properties given in the datasheet~\cite{CCF_Datasheet}. The rest of the cross-section is made up of PETG. The higher stiffness of the fiber composite material is accounted for using the transformed sections method \cite{Hibbeler_MoM}. 

The second moment of area of the beam is given by: 

\begin{equation}
    J_\text{beam} = \frac{b_\text{beam} \cdot h_\text{beam}^3 - b_\text{hollow} \cdot h_\text{hollow}^3}{12} +  2 n \left(\frac{b_\text{comp} \cdot h_\text{comp}^3}{12} + A_\text{comp} \cdot d_\text{NA}^2\right)
    \label{eq:MomentOfArea}
\end{equation}

\noindent where $n$ is the transformation factor, which is the ratio of stiffnesses of carbon fiber composite material and PETG: $n = \frac{E_\text{comp}}{E_\text{PETG}}$. $A_\text{comp}=h_\text{comp} \cdot b_\text{comp}$ is the area of the cross-section of the composite part.

\subsection{Three-point bending test}



In the linear regime of loading, the strain in the carbon fibers due to bending can be expressed as \cite{Hibbeler_MoM}:

\begin{equation}
\label{strain1}
    \epsilon = -\frac{d_\text{NA} \cdot M}{E_\text{PETG} \cdot J_\text{beam}}
\end{equation}

\noindent where $d_\text{NA}$ is the distance of the fibers from the neutral axis, and $M$ is the bending moment due to the loading.

The maximum deflection of the beam happens in the center, and can be expressed as~\cite{Hibbeler_MoM}:
\begin{equation}
\label{center_deflection}
    y_\text{max} = -\frac{P \cdot L^3}{48 \cdot E_\text{PETG} \cdot J_\text{beam}}
\end{equation}

Furthermore, the bending moment can be expressed as:

\begin{equation}
\label{bending moment}
    M = -\frac{P \cdot L}{4}
\end{equation}

By combining equations \ref{strain1}, \ref{center_deflection}, and \ref{bending moment}, the maximum strain in the fibers can be expressed as a function of constants and the center deflection:

\begin{equation}
    \epsilon_{\text{max}} = \frac{12 \cdot d_\text{NA} \cdot y_\text{max}}{L^2}
\end{equation}

Additionally, the average strain $\bar{\epsilon}$ in the fibers for the simply supported beam becomes $\bar{\epsilon} = \frac{1}{2} \epsilon_{\text{max}}$~\cite{Jonathan}.


The maximum loading stress during the three point bending test occurs in the fibers furthest from the neutral axis, and can be calculated as:
\begin{equation}
    {\sigma_\text{fiber,max}} = \frac{-M_\text{max} \cdot (d_\text{NA} + 0.5 \cdot h_\text{comp})} {J_\text{beam}} \cdot n
\end{equation}

\subsection{Residual Thermal Stress}
Due to the different coefficients of thermal expansion (CTE) of the carbon fibers and the PETG, there is significant residual thermal stress in the samples once the print has cooled, causing the carbon fibers to be in compression and the PETG to be in tension in the longitudinal direction. The composite and the PETG are modelled as two bars rigidly attached to each other, initially at a temperature higher than ambient temperature, but lower than the extrusion temperature due to rapid cooling after the extrusion \cite{delta_t_ambient}. It is assumed the temperature of the material is \SI{100}{\celsius} lower than the nozzle temperature by the time it gets deposited on the print, based on findings in \cite{delta_t_ambient}. The residual thermal stress on the fibers is:

\begin{equation}
\sigma_{\text{therm}} =\frac{E_{\text{PETG}} E_{\text{comp}} A_{\text{PETG}} \cdot (\alpha_{\text{PETG}} \Delta T_{\text{PETG}} - \alpha_{\text{fibers}} \Delta T_{\text{comp}})}{E_{\text{PETG}} A_{\text{PETG}} \left(1 + \alpha_{\text{fibers}} \Delta T_{\text{comp}}\right) + E_{\text{comp}} \cdot 2A_{\text{comp}} \left(1 + \alpha_{\text{PETG}} \Delta T_{\text{PETG}}\right)}
\end{equation}

\noindent where $\Delta T_\text{comp}$ and $\Delta T_\text{PETG}$ are the difference in temperature between the ambient environment and the hot materials when they are placed on the print: $\Delta T_\text{comp} = T_\text{ambient} - T_\text{comp}$, $\Delta T_\text{PETG} = T_\text{ambient} - T_\text{PETG}$. It is assumed the fibers dominate the thermal behaviour of the composite, due to their much higher Young's modulus and only slightly lower cross-sectional area.

\noindent The residual thermal stresses in the composite due to printing are compressive, and were calculated to be $\SI{-203}{\mega\Pa}$, \SI{-210}{\mega\Pa}, and \SI{-217}{\mega\Pa} for the Short, Medium, and Tall samples respectively. The parameters used to calculate these values are shown in table~\ref{tab:variables}, but it should be noted that both the model and the parameters are approximate, as described in more detail in the discussion.

\subsection{Fiber Strength}
Carbon fiber has a high tensile strength, but significantly lower compressive strength, which can be as little as 10 to 60 percent of the tensile strength~\cite{fiber_strength}. This puts the compressive yield strength of pure carbon fibers in the range of 1 to \SI{3}{\giga\pascal}, despite the tensile yield strength being in the range of 3 to \SI{7}{\giga\pascal} \cite{fiber_strength}. According to the CCF datasheet~\cite{CCF_Datasheet}, the composite material made up of the coextruded fibers with PETG has a tensile strength of \SI[separate-uncertainty = true]{774.4(27.1)}{\mega\pascal}, and a compressive strength of only \SI[separate-uncertainty = true]{237.4(4.2)}{\mega\pascal} when aligned with the load.

Given the high residual compressive thermal stress, and the high loading stress caused by three-point bending tests, fiber breakage due to compressive stress is realistic and fiber breakage due to tensile stress is much less likely.

\subsection{Electrical Model}
A linear model is used to model the electric properties of the fibers, as used in literature~\cite{Luan,Banks_review}. The change in resistance is assumed to be proportional to the average strain:

\begin{equation}
\label{dR}
    \Delta R = R_\text{0} \cdot k \cdot \bar{\epsilon}
\end{equation}

\noindent where $\Delta R$ is the change in resistance of the strained sample, $R_\text{0}$ is the initial (unstrained) resistance of the sample, $k$ is the strain sensitivity (or gauge factor), and $\bar{\epsilon}$ is the average strain in the fibers of the sample. The total resistance of the sample becomes:

\begin{equation}
    R = R_\text{0} \cdot (1 + k \cdot \bar{\epsilon})
\end{equation}

In some samples the fiber is co-extruded with Protopasta, theoretically affecting the resistance of the sample. However, due to the much higher resistivity of the Protopasta ($\rho_\text{Protopasta} = \SI{30}{\ohm\cm}$, $\rho_\text{fiber} = \SI{2}{\milli\ohm\cm}$) and a slightly higher cross-sectional area (at most 4 times higher), the resistance of the Protopasta section is over $3750$ higher than that of the carbon fibers, and is considered negligible.

\begin{table}[h!]
	\centering
     \footnotesize
	\begin{tabular}{m{0.1\textwidth} m{0.7\textwidth} m{0.2\textwidth}}
		\hline
		Parameter & Description & Value/Values\\
		\hline
		$d_\text{NA}$ & Distance from the neutral axis to the center of the extruded fibers & 2.1, 3.1, \SI{4.1}{\mm} \\
		 
		$h_\text{comp}$ & The height of the rectangle of coextruded fiber composite & \SI{0.78}{\mm}\\
		 
		$h_\text{beam}$ & The height of the beam & 6, 8, \SI{10}{\mm}\\
		 
		$h_\text{hollow}$ & The height of the hollow section in the center of the beam & 2.1, 4.1, \SI{6.1}{\mm}\\
		 
		$b_\text{comp}$ & The width of the rectangle of coextruded fiber composite & \SI{0.6}{\mm}\\
		 
		$b_\text{beam}$ & The width of the beam & \SI{9}{\mm}\\
		 
		$b_\text{hollow}$ & The width of the hollow section in the center of the beam & \SI{8.1}{\mm}\\

		 
		 

        $T_\text{ambient}$ & Ambient temperature. & \SI{20}{\celsius}\\

        $T_\text{comp}$ & Temperature of the extruded composite material when it gets placed on the print. & \SI{125}{\celsius} \cite{delta_t_ambient} \\

        $T_\text{PETG}$ & Temperature of the extruded PETG when it gets placed on the print. & \SI{140}{\celsius} \cite{delta_t_ambient} \\

        $\alpha_\text{fiber}$ & Coefficient of thermal expansion of the fibers & \SI{-1.0}{\micro\per\celsius} ~\cite{CTE_carbon_fiber}\\ 

        $\alpha_\text{PETG}$ & Coefficient of thermal expansion of PETG & \SI{50}{\micro\per\celsius} ~\cite{PETG_CTE}\\


        $E_\text{fiber}$ & Young's modulus of the carbon fibers & \SI{225}{\giga\Pa} ~\cite{fiber_strength} \\
        
		$E_\text{comp}$ & Young's modulus of the coextruded fiber composite & \SI{56.6}{\giga\Pa} ~\cite{CCF_Datasheet}\\
		 
		$E_\text{PETG}$ & Young's modulus of PETG & \SI{1.8}{\giga\Pa} \cite{PETG_TDS}\\

        $E_\text{PLA}$ & Young's modulus of PLA & \SI{2.4}{\giga\Pa} \cite{PLA_TDS}\\
		 
		\hline
	\end{tabular}
	\caption{Variables used in the theory section.}
	\label{tab:variables}
\end{table}
\section{Results}{\label{results}}

\subsection{Pre-stressing experiments}

\begin{figure}[h]
    \centering
    \includegraphics[width=\linewidth]{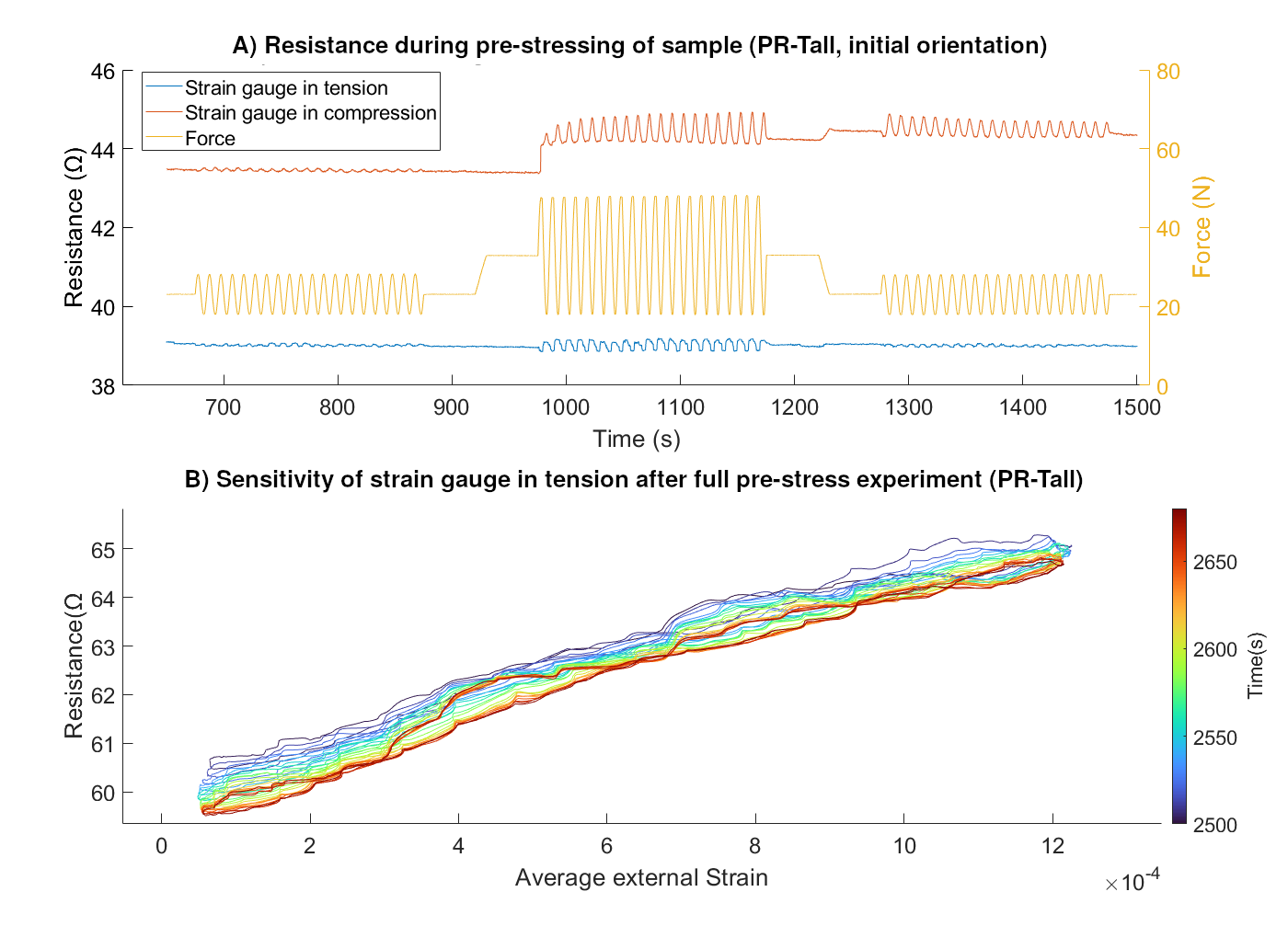}
    \caption{Experimental results of a Protopasta sample. A) Part of a pre-stress experiment, in which the force applied to the sample and the resistance of the strain gauges are plotted over time. The high amplitude loading causes a higher sensitivity even during low subsequent loading. B) Resistance against strain of a strain gauge in compression during low amplitude loading of the sample, after the sample has been fully pre-stressed. }
    \label{fig:results_pr}
\end{figure}


\FloatBarrier

The pre-stressing behaviour for the Protopasta samples can be seen in figure~\ref{fig:results_pr}A. When loaded with a small force, the sample shows almost no change in resistance (a sensitivity close to 0). When loaded with a sufficiently high force (\SI{48}{\newton} in this experiment), the sample starts to show a load-dependent change in resistance on the compressed side. This increase in sensitivity remains present when subsequently loaded with a small force. Hence, the high amplitude compressive loading caused an irreversible change in the strain gauge, increasing its sensitivity. 

\begin{figure}[h]
    \centering
    \includegraphics[width=\linewidth]{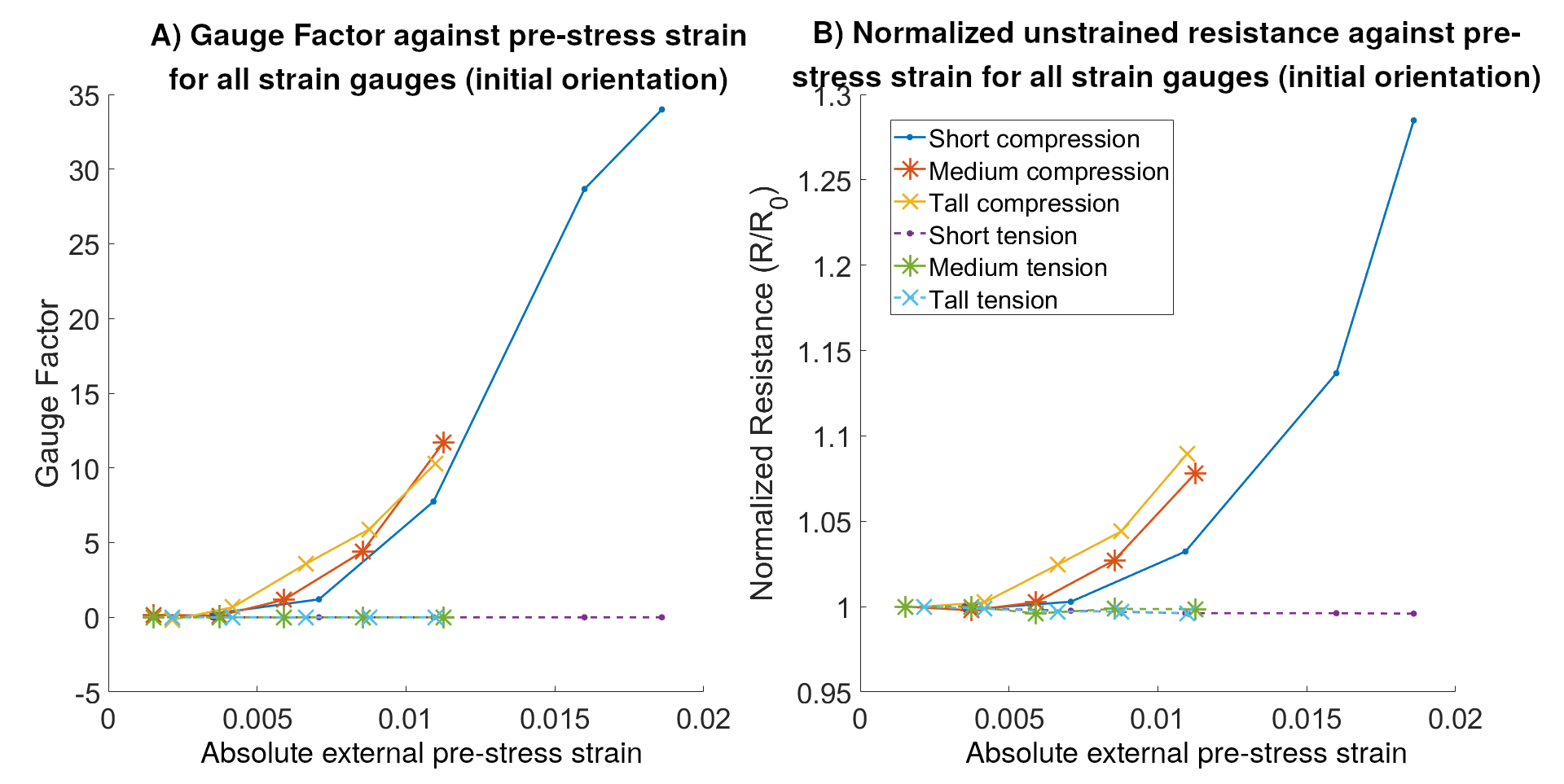}
    \caption{Gauge factor (A) and Resistance (B) against the maximum external strain experienced by the strain gauges during the pre-stress experiments in the initial orientation. }
    \label{fig:breakin_initial}
\end{figure}

For higher pre-stress loads, the sensitivity of the sample increases even more, figure~\ref{fig:breakin_initial}A. In the initial orientation, the sensitivities increase only for the strain gauges in compression, while the strain gauges in tension retain their gauge factor of approximately 0. The same trend can be seen for the unstrained resistance of the strain gauges in figure~\ref{fig:breakin_initial}B, with the resistance of the strain gauges in compression going up, and that of the strain gauges in tension not changing significantly.

\begin{figure}[h]
    \centering
    \includegraphics[width=\linewidth]{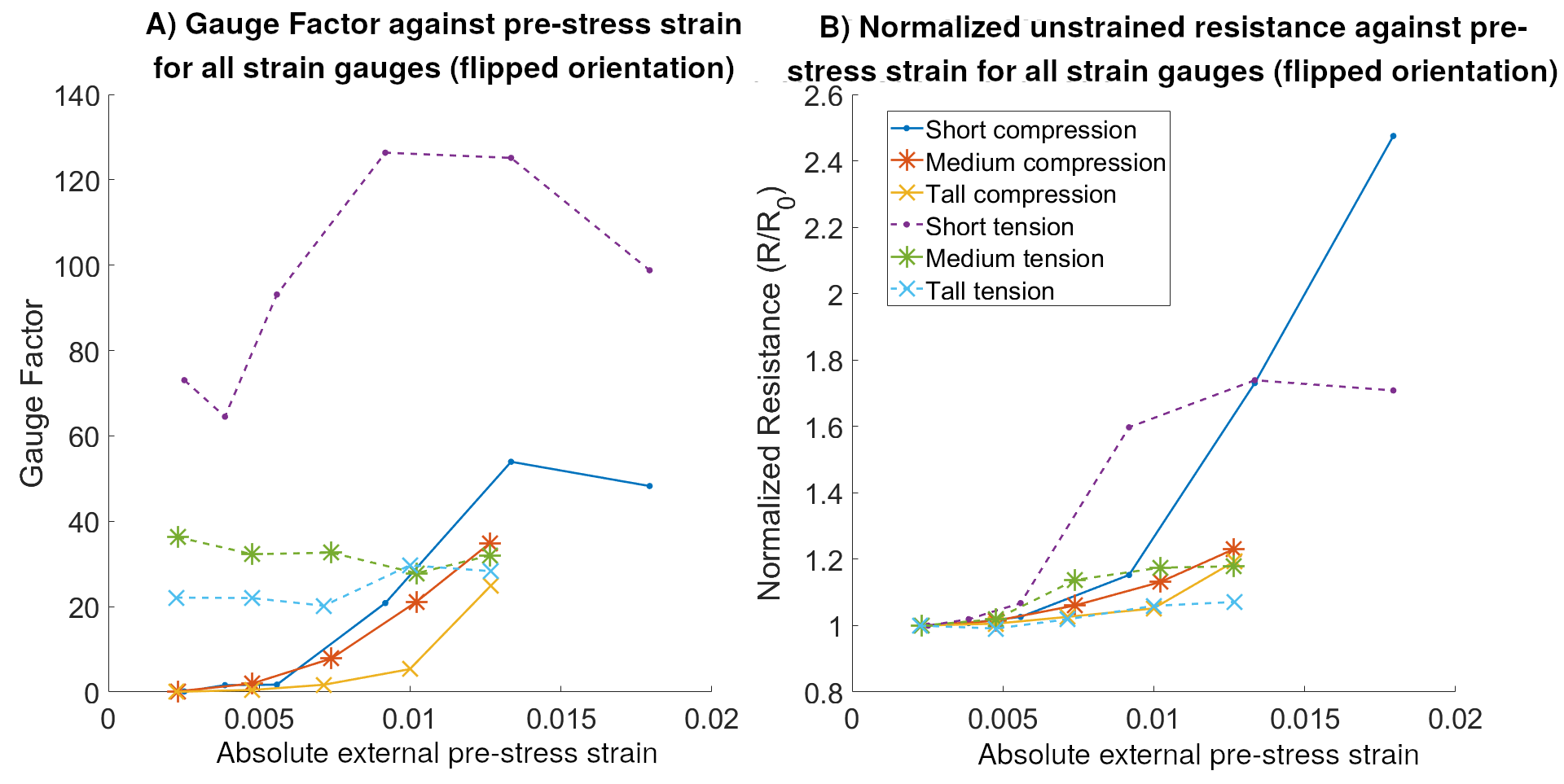}
    \caption{Gauge factor (A) and Resistance (B) against the maximum external  strain experienced by the strain gauges during the pre-stress experiments in the flipped orientation. }
    \label{fig:breakin_flipped}
\end{figure}

In the flipped orientation, figure~\ref{fig:breakin_flipped}A, the strain gauges in compression follow the same trend as for the initial orientation: their gauge factor increases. It can be seen however that the gauge factor increases more than for the initial orientation. Furthermore, the strain gauges in tension also become more sensitive, with the Short sample reaching a gauge factor of 126, a lot higher than the standard values achieved in literature~\cite{new_piezo,Iizuka,fracture_detection}. When loaded sufficiently high, the gauge factor then starts decreasing again. The resistances also increase due to the loading strain, figure~\ref{fig:breakin_flipped}B.

The change in resistance increases with the increasing strain experienced by the carbon fibers, figure~\ref{fig:results_pr}B. It can be seen that the response follows a slightly sublinear trend, with minor hysteresis. Another interesting result that can be seen in figure~\ref{fig:results_pr}A is that when testing the sensitivity after pre-stressing, the first peak in the resistance response is the highest, with the subsequent peaks being slightly lower. The resistance of the sample itself also slightly decreases over time with continued cyclic loading, figures~\ref{fig:results_pr} A and B, which has also been observed in~\cite{new_piezo, Iizuka}.

The samples printed with PETG exhibit the same changes after pre-stressing as the samples printed with Protopasta. However, there is a significant difference in the level of noise and reliability of the samples. Out of the six strain gauges printed with Protopasta, all of them had good and reliable electrical contact with the fibers, as seen from the noise-free resistance measurements in figure~\ref{fig:results_pr}. From the six strain gauges printed with PETG, several issues were present: for two of them, the contacts failed to properly connect with the fibers, causing a contact resistance too high to measure useful data. From the remaining four, three of them showed high levels of noise, and only one showed an unambiguous response with little noise.

\begin{figure}[H]
\label{fig:PETG_break_in}
\includegraphics[width=\linewidth]{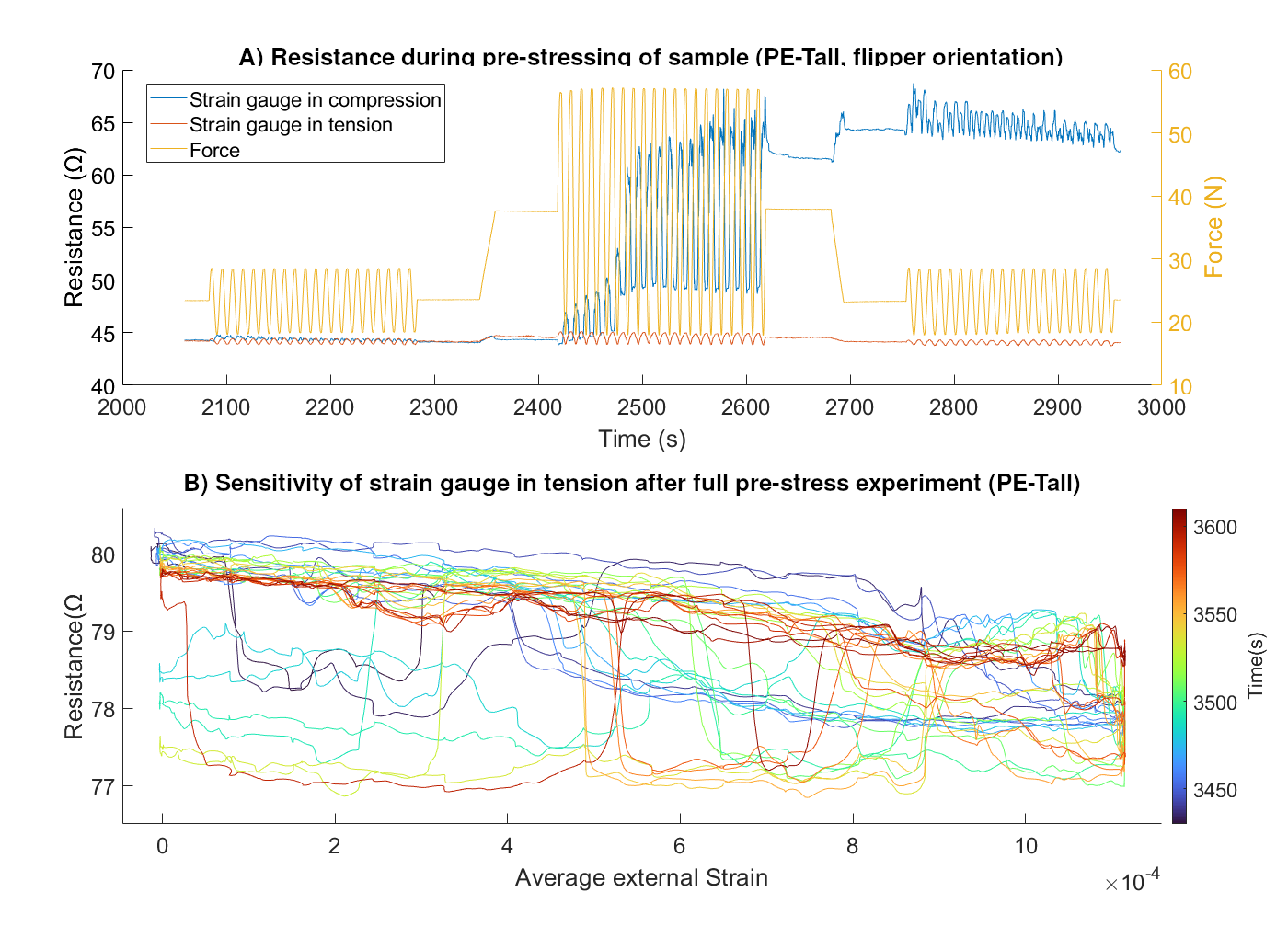}
\caption{Experimental results of a PETG sample. A) Part of a pre-stress experiment, in which the force applied to the sample and the resistance of the strain gauges are plotted over time. The high amplitude loading causes a higher resistance response even during low subsequent loading. This PETG sample exhibits very high noise, unlike the Protopasta samples. B) Resistance against strain of a strain gauge in compression during low amplitude loading of the sample, after the sample has been fully pre-stressed. }
\label{fig:results_pe}
\end{figure}

A part of the pre-stress experiments of a PETG sample can be seen in figure~\ref{fig:results_pe} A. It exhibits the same behaviour as the Protopasta samples: high loading causes a permanent change in the resistance and sensitivity of the sample. One notable difference is the noise that this sample exhibits. The jump in resistance is also a lot higher compared to the increase in resistance in the Protopasta samples, suggesting the effect of fibers breaking is larger in the PETG, because of its inability to bridge the connection between the fibers, unlike the Protopasta. After the complete pre-stress experiment, the response of the sample remains noisy, figure~\ref{fig:results_pe} B. It should be noted that one PETG sample showed a clean response, so it is possible to have quality samples made without Protopasta, but they are less likely to work, and more susceptible to noise.

Overall, printing with Protopasta instead of PETG has shown several benefits, including more reliable contact with the fibers, less noise in the electrical response of the samples, and less hysteresis in the final resistance-strain response of the pre-stressed samples. No negative effects have been observed in the performance of the samples. However, the nozzle was observed to clog more often while printing with Protopasta.

After pre-stressing the samples exhibit a high gauge factor (sensitivity), table~\ref{tab:sensitivities}, showing high potential for use as a sensor. Depending on how much they were pre-stressed and the size of the sample, the strain gauges also vary in terms of how linear the resistance response to strain is.

\begin{table}[h]
\footnotesize
\begin{tabular}{m{2cm} m{3cm} m{3cm} m{2.5cm} m{2.5cm}}
\textbf{Height (\SI{}{\mm}) } & \textbf{Max sensitivity in compression} & \textbf{Max sensitivity in tension} & \textbf{$R^2$ compression} & \textbf{$R^2$ tension} \\
\hline
6              & 53.94                                   & 126                                 & 0.918                   & 0.970                \\
8                  & 34.83                                   & 36.23                               & 0.923                   & 0.902               \\
10                 & 24.84                                   & 20.91                               & 0.868                   & 0.889              
\end{tabular}
\caption{Maximum sensitivities of the samples printed with Protopasta after the pre-stress experiment, accompanied by the coefficients of determination of the linear fit of $\Delta R$ and $\epsilon$ relationship predicted by the electrical model.}
\label{tab:sensitivities}
\end{table}

The samples experience high stress during the pre-stress experiments, table~\ref{tab:stresses}. The smaller samples experience more stress, as expected. It can also be seen that for all samples, the stress is higher than the fibers' compression yield stress of $237.4 \pm 4.2 \si{\mega\Pa}$ and lower than the fibers' tension yield stress of $774.4 \pm 27.1 \si{\mega\Pa}$. Thus, yielding of the fibers is expected to happen in compression, but not in tension.

\begin{table}[h]
\begin{tabular}{m{2cm} m{3cm} m{3cm} m{5cm} m{2cm}}
\textbf{Height (\SI{}{\mm})} & \textbf{Max loading stress (\SI{}{\mega\Pa})} & \textbf{Residual thermal stress (\SI{}{\mega\Pa})} & \textbf{Total compressive stress (\SI{}{\mega\Pa})}\\
\hline
6     & -488  & -203    & -691 \\
8     & -351	& -210    & -561 \\
10    & -251  & -217    & -468 \\            
\end{tabular}

\caption{Maximum loading stress, residual thermal stress, and total estimated compressive stress (loading $+$ residual thermal stress) of the samples printed with Protopasta.}
\label{tab:stresses}
\end{table}


\subsection{Micrograph of a sample}

\begin{figure}[H]
    \centering
    \includegraphics[width=0.9\textwidth]{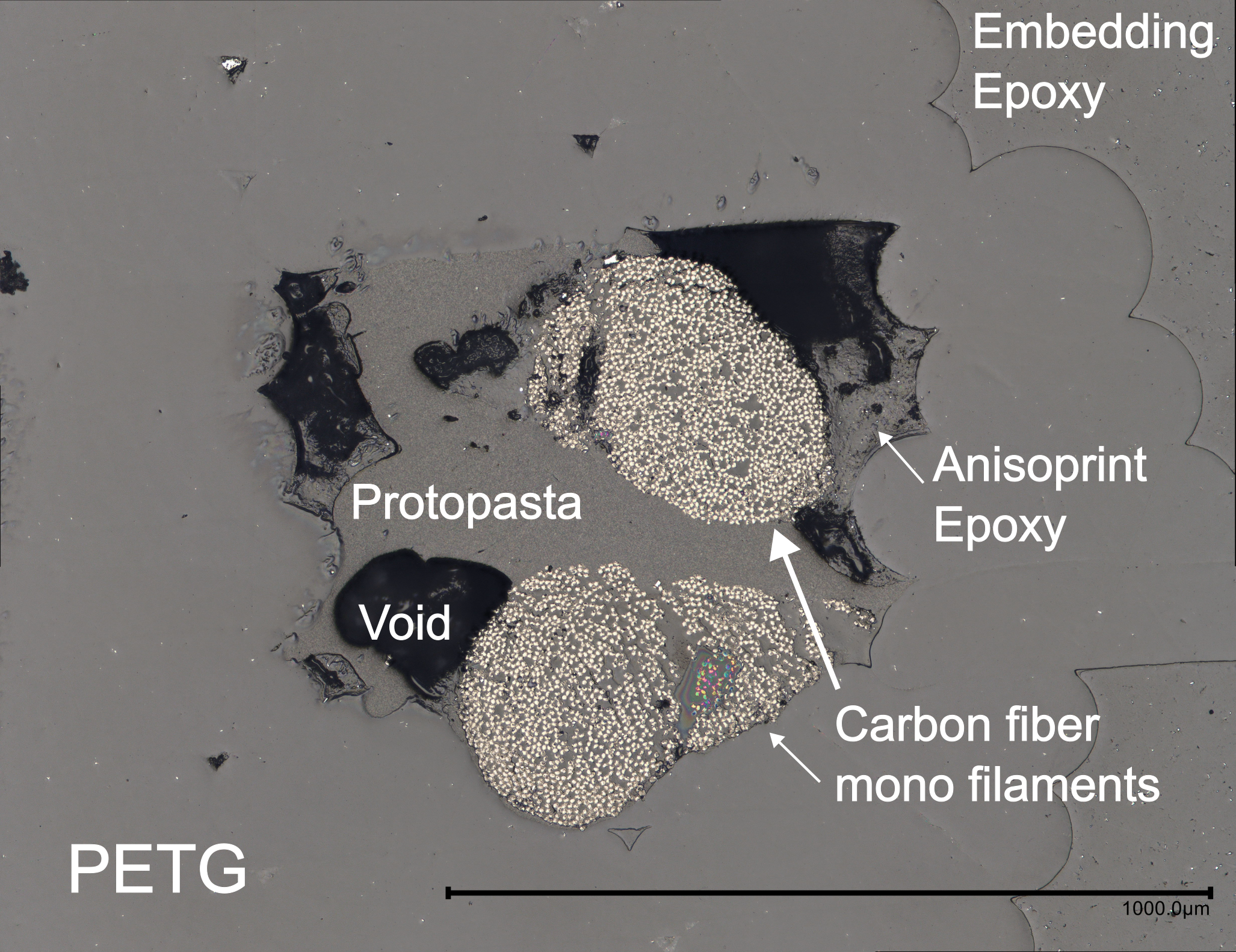}
    \caption{Micrograph of the cross-section of one sample printed with two CCF filaments.}
    \label{fig:micrograph}
\end{figure}

Figure~\ref{fig:micrograph} shows a micrograph of a sample printed with Protopasta and two CCF filaments.

Several things can be observed from the micrograph: for one, there are significant voids in the printed structure. Another important observation is the fact that the Anisoprint epoxy resin does not fully encompass the fibers after printing. This allows the Protopasta to contact the fibers, establishing an electrical contact between the fibers and Protopasta. It also means that many of the fibers are now (also) connected to each other through the Protopasta. Furthermore, shearing of the fiber bundles can be observed, as well as voids in the print, which shows that the printing process results in a number of defects.

\section{Discussion}{\label{discussion}}

\subsection{Mechanism for increase in sensitivity}
Taking into account the results and existing literature~\cite{Chang2025-he,LIU2025}, the likely dominant mechanism for the change in sensitivity during loading is fiber breakage. When loaded sufficiently, in a fiber some of the carbon monofilaments break due to the compressive stress. When strained, these broken monofilaments lose contact with each other, causing an increase in the strain gauge resistance. More broken monofilaments means an increase in electrical contact loss, causing a higher sensitivity. This is likely what was observed in the results, since more pre-stress load caused a higher increase in sensitivity. When too many monofilaments have broken, some monofilaments lose contact entirely (pull-out), which might cause the decrease in sensitivity observed in figure~\ref{fig:breakin_flipped} A. Before the decrease in sensitivity, the gauge factors achieved by the strain gauges are a lot higher than what is possible with the intrinsic properties of carbon fiber~\cite{CF_piezoresistive}, which could indicate that these strain gauges operate similarly to crack-based sensors~\cite{zhou2022}. 

Given the high stresses experienced by the fibers due to loading and the residual thermal stresses, table~\ref{tab:stresses}, fiber breakage is solely expected in compression. As seen in figure~\ref{fig:results_pr}, the sensitivity only increases for the samples in compression, which indicates change in sensitivity and fiber breakage are related. It should be noted that the calculated stress values only apply in the linear region of the material, before any of the fibers break, so the real stresses after the first fibers break are different than calculated. This explains why in the flipped orientation, the sensitivity increases for the fibers in tension too; the stress is higher than calculated due to a damage-induced local stress concentration at the place where the fibers broke in the initial orientation.  Fiber breakage and other failure mechanisms, such as fiber/matrix debonding and matrix cracking, combined with the pre-existing voids, will lead to higher stress concentrations and hence higher strain on the fibers, ultimately leading to higher gauge factors.


Additionally, the increase in the absolute resistance of the strain gauge, figures \ref{fig:results_pr} B and \ref{fig:breakin_flipped} B, suggests that fibers break. Lastly, crackling noises could be heard during the pre-stress process, which is likely due to fiber failure. The audible failure of carbon fibers was also observed in~\cite{new_piezo}.

Fiber breakage causing an increase in sensitivity has also been mentioned in literature~\cite{new_piezo,Iizuka,Yao}, with the increase in the sensitivity before total failure usually attributed to progressive fiber fracture. The pre-stress load has only been applied in the middle of the beam. In future work it is interesting to study the location dependence of the pre-stressing on the sensitivity increase.

\subsection{Effects of Conductive Filament}
The big reduction of noise caused by the Protopasta is most likely due to its conductivity, which allows alternative current paths parallel to the carbon fibers. The noise in the samples, which is also present in literature~\cite{Jonathan,Luan,Yao}, could potentially be mitigated by co-extruding conductive filament with the carbon fibers. It is still unclear to what extent the positive effect of the conductive filament is due to its ability to electrically connect broken fibers, or due to its ability to connect the fibers to the electrodes. Regardless, the effects of Protopasta on the performance of the strain gauge are positive. The positive effect of conductive material has also been shown in crack-based sensors, allowing the effective management of the electrical performance of the sensor~\cite{zhou2022}.

\subsection{Drift in resistance}
Drift is present in the resistance response, figure~\ref{fig:results_pr} B and figure~\ref{fig:results_pe} B, as was also  observed in literature~\cite{new_piezo,Iizuka}. Iizuka and Todoroki~\cite{Iizuka} credit the decrease in resistance over multiple cycles to viscoelastic effects in the matrix surrounding the carbon fibers. The drift is quite consistent and relatively small between cycles, figure~\ref{fig:results_pr} B. High pass filtering can likely be used for future sensing applications. It should be noted that a different rate of pre-stressing also influences the achieved sensitivity~\cite{LIU2025}.


\subsection{Limitations and suggested improvements}
The maximum force of the linear actuator used to load the samples was a setup limitation. In order to increase this maximum force, attaching weights was necessary, resulting in a limited achievable force range that could be applied to the samples (\SI{20}{\N} to \SI{70}{\N}). Due to the constant positive force on the sample, each strain gauge could be either in compression or in tension, but it was not possible to do both within the same experiment. Furthermore, the force was never \SI{0}{\newton} which possibly caused viscoelastic deformation over the long course of the experiments. Additionally, due to a homing error with the linear actuator, there is an unknown offset in the center deflection data of the samples, which makes determining the absolute strain challenging. For this reason, the relative (peak to peak) strain is used for plotting and the reported external pre-stress strain is lower than the actual strain.

Another limitation of this research is the use of the two electrodes for the measurement of the resistance instead of a four probe set-up. This caused the measured gauge factor to be under reported due to the inclusion of the contact resistance~\cite{Chung}. However, given the very large gauge factors measured and the fact that the contacts were unstrained in this experiment, this is not a very significant limitation. The reason two contacts were used was due to the difficulty of establishing four contact probes. For one, the Anisoprint Composer can only print fibers on the inside walls of the printed structure, so the surrounding PETG would need to be melted to reach the fibers. Then, the brittle fibers would need to be contacted without breaking them, which is a challenging task. Overall, for the goals of this research, the two probe method was sufficient, but future research could implement a four probe method in order to get more accurate resistance readings.

Additionally, due to the fabrication process, some difficulties arise in the theoretical analysis of the beams. For one, there are many factors that affect the properties and cross-section of the materials, such as temperature, extrusion speed, the nozzle diameter, the layer height, the presence of voids, and more. Because of this, the values for stiffness and ultimate strength of the printing materials cannot be known exactly. To add to this, the Anisoprint datasheet~\cite{CCF_Datasheet} provides data for a tensile test, and not a flexural test. Furthermore, during the printing process, the material is heated and subsequently cools down, introducing residual thermal stresses in the samples. Modelling this process is complex, and beyond the scope of this paper, so only a simplified model was used. Additionally, the temperature difference between the printing temperature and the ambient temperature is not trivial. The temperature difference is a rough approximation, since the material significantly cools down from its extrusion temperature by the time it solidifies. Based on the high calculated values (\SI{-203}{}  to \SI{-217}{\mega\Pa}), the model likely overestimated the residual thermal stress, since these values come close to the yield stress in compression. 

Lastly, the fiber breakage could not be shown on the micrograph, due to only being able to make transverse cuts of the sample. A micrograph of a longitudinal cut or micro-CT could potentially give definite proof of fiber breakage, as shown in~\cite{Chang2025-he}. 

\subsection{Potential for use of 3D printed carbon fiber based strain gauges}
There is significant potential for the use of carbon fiber as sensors. One significant advantage is the very high gauge factors (over $100$) than can be achieved with the pre-stressing method, making the sensor very sensitive. Another advantage is the relative ease of fabrication, which only requires an Anisoprint 3D printer and cheap tools (saw, screws, silver paint, screw driver). Finally, the load-bearing carbon fibers measure their own strain inside the structure, unlike standard adhesive strain gauges which are limited to measuring strain on the surface. 

The drift caused by viscoelastic effects is a disadvantage, affecting the sensor precision and accuracy. Furthermore, using the method in this paper requires significant pre-stressing in order to achieve the high gauge factors, likely causing fiber fractures. This might not be feasible for all critical structures, and also affects the ultimate strength of the structure. The structure will still benefit from the stiffness and strength of the fibers, though its strength will be reduced. 



\section{Conclusion}{\label{conclusion}}

In this paper, highly sensitive carbon fiber strain gauges were 3D printed and made more sensitive by pre-stressing the samples under high loads, achieving gauge factors of over $100$. It was shown that the increase in gauge factor at high loads is an irreversible change that also persists at low loading of the fibers, offering great potential for sensing applications. The increase in gauge factor is attributed to progressive fiber failure. The coextrusion of conductive filament alongside the fibers was investigated, and the results showed that it reduces the noise in the samples, and improves the electrical connection between the electrode and the carbon fibers. More research needs to be done in order to make a full characterization of the sensors' properties. 

\section*{CRediT authorship contribution statement}
\textbf{Matei Drilea}: Conceptualization, Formal analysis, Investigation, Methodology, Writing – original draft, Writing – review \& editing. \textbf{Alexander Dijkshoorn}: Conceptualization, Formal analysis, Methodology, Resources, Writing – original draft, Writing – review \& editing. \textbf{Gusthavo Ribeiro Salom\~{a}o}: Investigation, Resources, Writing – review \& editing. \textbf{Stefano Stramigioli}: Funding acquisition, Supervision, Writing - review \& editing. \textbf{Gijs Krijnen}: Conceptualization, Formal analysis, Project administration, Supervision, Writing – review \& editing. 

\section*{Declaration of competing interest}
The authors declare that they have no known competing financial interests or personal relationships that could have appeared to influence the work reported in this paper.

\section*{Data availability}
Data will be made available on request.

\section*{Acknowledgements}
This work was partially supported by the PortWings Project, funded by the European Research Council under Agreement 787675.

\bibliographystyle{elsarticle-num-names}
\label{ch:bib} 
\bibliography{bibliography} 

\begin{thebibliography}{37}
\expandafter\ifx\csname natexlab\endcsname\relax\def\natexlab#1{#1}\fi
\providecommand{\url}[1]{\texttt{#1}}
\providecommand{\href}[2]{#2}
\providecommand{\path}[1]{#1}
\providecommand{\DOIprefix}{doi:}
\providecommand{\ArXivprefix}{arXiv:}
\providecommand{\URLprefix}{URL: }
\providecommand{\Pubmedprefix}{pmid:}
\providecommand{\doi}[1]{\href{http://dx.doi.org/#1}{\path{#1}}}
\providecommand{\Pubmed}[1]{\href{pmid:#1}{\path{#1}}}
\providecommand{\bibinfo}[2]{#2}
\ifx\xfnm\relax \def\xfnm[#1]{\unskip,\space#1}\fi
\bibitem[{Molitch-Hou(2018)}]{AM_intro}
\bibinfo{author}{M.~Molitch-Hou},
\newblock \bibinfo{title}{1 - overview of additive manufacturing process},
\newblock in: \bibinfo{editor}{J.~Zhang}, \bibinfo{editor}{Y.-G. Jung} (Eds.), \bibinfo{booktitle}{Additive Manufacturing}, \bibinfo{publisher}{Butterworth-Heinemann}, \bibinfo{year}{2018}, pp. \bibinfo{pages}{1--38}. \URLprefix \url{https://www.sciencedirect.com/science/article/pii/B9780128121559000013}. \DOIprefix\doi{https://doi.org/10.1016/B978-0-12-812155-9.00001-3}.
\bibitem[{Blanco(2020)}]{AM_weak}
\bibinfo{author}{I.~Blanco},
\newblock \bibinfo{title}{The use of composite materials in 3d printing},
\newblock \bibinfo{journal}{Journal of Composites Science} \bibinfo{volume}{4} (\bibinfo{year}{2020}). \URLprefix \url{https://www.mdpi.com/2504-477X/4/2/42}. \DOIprefix\doi{10.3390/jcs4020042}.
\bibitem[{Parandoush and Lin(2017)}]{CFR_usage}
\bibinfo{author}{P.~Parandoush}, \bibinfo{author}{D.~Lin},
\newblock \bibinfo{title}{A review on additive manufacturing of polymer-fiber composites},
\newblock \bibinfo{journal}{Composite Structures} \bibinfo{volume}{182} (\bibinfo{year}{2017}) \bibinfo{pages}{36--53}. \URLprefix \url{https://www.sciencedirect.com/science/article/pii/S0263822316329063}. \DOIprefix\doi{https://doi.org/10.1016/j.compstruct.2017.08.088}.
\bibitem[{Zhuo et~al.(2021)Zhuo, Li, Ashcroft, and Jones}]{newer_review}
\bibinfo{author}{P.~Zhuo}, \bibinfo{author}{S.~Li}, \bibinfo{author}{I.~A. Ashcroft}, \bibinfo{author}{A.~I. Jones},
\newblock \bibinfo{title}{Material extrusion additive manufacturing of continuous fibre reinforced polymer matrix composites: A review and outlook},
\newblock \bibinfo{journal}{Composites Part B: Engineering} \bibinfo{volume}{224} (\bibinfo{year}{2021}) \bibinfo{pages}{109143}. \URLprefix \url{https://www.sciencedirect.com/science/article/pii/S1359836821005266}. \DOIprefix\doi{https://doi.org/10.1016/j.compositesb.2021.109143}.
\bibitem[{Cheng et~al.(2023)Cheng, Peng, Li, Rao, {Le Duigou}, Wang, and Ahzi}]{CHENG2023}
\bibinfo{author}{P.~Cheng}, \bibinfo{author}{Y.~Peng}, \bibinfo{author}{S.~Li}, \bibinfo{author}{Y.~Rao}, \bibinfo{author}{A.~{Le Duigou}}, \bibinfo{author}{K.~Wang}, \bibinfo{author}{S.~Ahzi},
\newblock \bibinfo{title}{3d printed continuous fiber reinforced composite lightweight structures: A review and outlook},
\newblock \bibinfo{journal}{Composites Part B: Engineering} \bibinfo{volume}{250} (\bibinfo{year}{2023}) \bibinfo{pages}{110450}. \DOIprefix\doi{https://doi.org/10.1016/j.compositesb.2022.110450}.
\bibitem[{Karaboğa et~al.(2024)Karaboğa, Göleç, Yunus, Toros, and Öz}]{CF_aero}
\bibinfo{author}{F.~Karaboğa}, \bibinfo{author}{F.~Göleç}, \bibinfo{author}{D.~E. Yunus}, \bibinfo{author}{S.~Toros}, \bibinfo{author}{Y.~Öz},
\newblock \bibinfo{title}{Mechanical response of carbon fiber reinforced epoxy composite parts joined with varying bonding techniques for aerospace applications},
\newblock \bibinfo{journal}{Composite Structures} \bibinfo{volume}{331} (\bibinfo{year}{2024}) \bibinfo{pages}{117920}. \URLprefix \url{https://www.sciencedirect.com/science/article/pii/S0263822324000485}. \DOIprefix\doi{https://doi.org/10.1016/j.compstruct.2024.117920}.
\bibitem[{Gardie et~al.(2021)Gardie, Paramasivam, Dubale, {Tefera Chekol}, and Selvaraj}]{CF_automotive}
\bibinfo{author}{E.~Gardie}, \bibinfo{author}{V.~Paramasivam}, \bibinfo{author}{H.~Dubale}, \bibinfo{author}{E.~{Tefera Chekol}}, \bibinfo{author}{S.~K. Selvaraj},
\newblock \bibinfo{title}{Numerical analysis of reinforced carbon fiber composite material for lightweight automotive wheel application},
\newblock \bibinfo{journal}{Materials Today: Proceedings} \bibinfo{volume}{46} (\bibinfo{year}{2021}) \bibinfo{pages}{7369--7374}. \URLprefix \url{https://www.sciencedirect.com/science/article/pii/S2214785320407655}. \DOIprefix\doi{https://doi.org/10.1016/j.matpr.2020.12.1047}, \bibinfo{note}{3rd International Conference on Materials, Manufacturing and Modelling}.
\bibitem[{Zhang et~al.(2021)Zhang, Tian, Raza, Zhao, Wang, Du, Zhang, and Qu}]{CF_robotics}
\bibinfo{author}{X.~Zhang}, \bibinfo{author}{M.~Tian}, \bibinfo{author}{T.~Raza}, \bibinfo{author}{H.~Zhao}, \bibinfo{author}{J.~Wang}, \bibinfo{author}{X.~Du}, \bibinfo{author}{X.~Zhang}, \bibinfo{author}{L.~Qu},
\newblock \bibinfo{title}{Soft robotic reinforced by carbon fiber skeleton with large deformation and enhanced blocking forces},
\newblock \bibinfo{journal}{Composites Part B: Engineering} \bibinfo{volume}{223} (\bibinfo{year}{2021}) \bibinfo{pages}{109099}. \URLprefix \url{https://www.sciencedirect.com/science/article/pii/S1359836821004832}. \DOIprefix\doi{https://doi.org/10.1016/j.compositesb.2021.109099}.
\bibitem[{{Anisoprint Support}(2023)}]{Composer_Manual}
\bibinfo{author}{{Anisoprint Support}}, \bibinfo{title}{Composer user manual}, \bibinfo{year}{2023}. \URLprefix \url{https://support.anisoprint.com/composer/manual/}, \bibinfo{note}{accessed at 2024-03-12}.
\bibitem[{Azarov et~al.(2019)Azarov, Antonov, Golubev, Khaziev, and Ushanov}]{Anisoprint_paper}
\bibinfo{author}{A.~V. Azarov}, \bibinfo{author}{F.~K. Antonov}, \bibinfo{author}{M.~V. Golubev}, \bibinfo{author}{A.~R. Khaziev}, \bibinfo{author}{S.~A. Ushanov},
\newblock \bibinfo{title}{Composite 3d printing for the small size unmanned aerial vehicle structure},
\newblock \bibinfo{journal}{Composites Part B: Engineering} \bibinfo{volume}{169} (\bibinfo{year}{2019}) \bibinfo{pages}{157--163}. \URLprefix \url{https://www.sciencedirect.com/science/article/pii/S1359836818320031}. \DOIprefix\doi{https://doi.org/10.1016/j.compositesb.2019.03.073}.
\bibitem[{Scholle and Sinapius(2021)}]{CF_piezoresistive}
\bibinfo{author}{P.~Scholle}, \bibinfo{author}{M.~Sinapius},
\newblock \bibinfo{title}{A review on the usage of continuous carbon fibers for piezoresistive self strain sensing fiber reinforced plastics},
\newblock \bibinfo{journal}{Journal of Composites Science} \bibinfo{volume}{5} (\bibinfo{year}{2021}). \URLprefix \url{https://www.mdpi.com/2504-477X/5/4/96}. \DOIprefix\doi{10.3390/jcs5040096}.
\bibitem[{Chung(2020)}]{Chung}
\bibinfo{author}{D.~Chung},
\newblock \bibinfo{title}{A critical review of piezoresistivity and its application in electrical-resistance-based strain sensing},
\newblock \bibinfo{journal}{Journal of Materials Science}  (\bibinfo{year}{2020}). \URLprefix \url{https://doi.org/10.1007s10853-020-05099-z}. \DOIprefix\doi{10.1007/s10853-020-05099-z}.
\bibitem[{{Wan A Hamid} et~al.(2023){Wan A Hamid}, Iannucci, and Robinson}]{CF_flexural}
\bibinfo{author}{W.~{Wan A Hamid}}, \bibinfo{author}{L.~Iannucci}, \bibinfo{author}{P.~Robinson},
\newblock \bibinfo{title}{Flexural behaviour of 3d-printed carbon fibre composites: Experimental and virtual tests - application to composite adaptive structure},
\newblock \bibinfo{journal}{Composites Part C: Open Access} \bibinfo{volume}{10} (\bibinfo{year}{2023}) \bibinfo{pages}{100344}. \URLprefix \url{https://www.sciencedirect.com/science/article/pii/S2666682022001074}. \DOIprefix\doi{https://doi.org/10.1016/j.jcomc.2022.100344}.
\bibitem[{Parker et~al.(2023)Parker, Ezeokeke, Matsuzaki, and Arola}]{CFR_printing_weaken}
\bibinfo{author}{M.~Parker}, \bibinfo{author}{N.~Ezeokeke}, \bibinfo{author}{R.~Matsuzaki}, \bibinfo{author}{D.~Arola},
\newblock \bibinfo{title}{Strength and its variability in 3d printing of polymer composites with continuous fibers},
\newblock \bibinfo{journal}{Materials \& Design} \bibinfo{volume}{225} (\bibinfo{year}{2023}) \bibinfo{pages}{111505}. \URLprefix \url{https://www.sciencedirect.com/science/article/pii/S0264127522011285}. \DOIprefix\doi{https://doi.org/10.1016/j.matdes.2022.111505}.
\bibitem[{Schaaij et~al.(2023)Schaaij, Dijkshoorn, Stramigioli, and Krijnen}]{Jonathan}
\bibinfo{author}{J.~Schaaij}, \bibinfo{author}{A.~Dijkshoorn}, \bibinfo{author}{S.~Stramigioli}, \bibinfo{author}{G.~Krijnen},
\newblock \bibinfo{title}{Self-sensing properties of continuous carbon fiber reinforced, 3d-printed beams},
\newblock in: \bibinfo{booktitle}{SMSI 2023 Conference}, \bibinfo{year}{2023}. \URLprefix \url{https://www.smsi-conference.com/}. \DOIprefix\doi{10.5162/SMSI2023/P63}, \bibinfo{note}{sensor and Measurement Science International Conference, SMSI 2023 , SMSI 2023germ ; Conference date: 08-05-2023 Through 11-05-2023}.
\bibitem[{Luan et~al.(2018)Luan, Yao, Shen, and Fu}]{Luan}
\bibinfo{author}{C.~Luan}, \bibinfo{author}{X.~Yao}, \bibinfo{author}{H.~Shen}, \bibinfo{author}{J.~Fu},
\newblock \bibinfo{title}{Self-sensing of position-related loads in continuous carbon fibers-embedded 3d-printed polymer structures using electrical resistance measurement},
\newblock \bibinfo{journal}{Sensors} \bibinfo{volume}{18} (\bibinfo{year}{2018}). \URLprefix \url{https://www.mdpi.com/1424-8220/18/4/994}. \DOIprefix\doi{10.3390/s18040994}.
\bibitem[{Heitkamp et~al.(2024)Heitkamp, Goutier, Hilbig, Girnth, Waldt, Klawitter, and Vietor}]{new_piezo}
\bibinfo{author}{T.~Heitkamp}, \bibinfo{author}{M.~Goutier}, \bibinfo{author}{K.~Hilbig}, \bibinfo{author}{S.~Girnth}, \bibinfo{author}{N.~Waldt}, \bibinfo{author}{G.~Klawitter}, \bibinfo{author}{T.~Vietor},
\newblock \bibinfo{title}{Parametric study of piezoresistive structures in continuous fiber reinforced additive manufacturing},
\newblock \bibinfo{journal}{Composites Part C: Open Access} \bibinfo{volume}{13} (\bibinfo{year}{2024}) \bibinfo{pages}{100431}. \URLprefix \url{https://www.sciencedirect.com/science/article/pii/S2666682024000021}. \DOIprefix\doi{https://doi.org/10.1016/j.jcomc.2024.100431}.
\bibitem[{Yao et~al.(2017)Yao, Luan, Zhang, Lan, and Fu}]{Yao}
\bibinfo{author}{X.~Yao}, \bibinfo{author}{C.~Luan}, \bibinfo{author}{D.~Zhang}, \bibinfo{author}{L.~Lan}, \bibinfo{author}{J.~Fu},
\newblock \bibinfo{title}{Evaluation of carbon fiber-embedded 3d printed structures for strengthening and structural-health monitoring},
\newblock \bibinfo{journal}{Materials \& Design} \bibinfo{volume}{114} (\bibinfo{year}{2017}) \bibinfo{pages}{424--432}. \URLprefix \url{https://www.sciencedirect.com/science/article/pii/S0264127516313879}. \DOIprefix\doi{https://doi.org/10.1016/j.matdes.2016.10.078}.
\bibitem[{Billah et~al.(2021)Billah, Coronel, Chavez, Lin, and Espalin}]{Billah}
\bibinfo{author}{K.~M.~M. Billah}, \bibinfo{author}{J.~L. Coronel}, \bibinfo{author}{L.~Chavez}, \bibinfo{author}{Y.~Lin}, \bibinfo{author}{D.~Espalin},
\newblock \bibinfo{title}{Additive manufacturing of multimaterial and multifunctional structures via ultrasonic embedding of continuous carbon fiber},
\newblock \bibinfo{journal}{Composites Part C: Open Access} \bibinfo{volume}{5} (\bibinfo{year}{2021}) \bibinfo{pages}{100149}. \URLprefix \url{https://www.sciencedirect.com/science/article/pii/S266668202100044X}. \DOIprefix\doi{https://doi.org/10.1016/j.jcomc.2021.100149}.
\bibitem[{Chang et~al.(2025)Chang, Zhang, Gu, Fu, and Han}]{Chang2025-he}
\bibinfo{author}{C.~Chang}, \bibinfo{author}{P.~Zhang}, \bibinfo{author}{J.~Gu}, \bibinfo{author}{H.~Fu}, \bibinfo{author}{Z.~Han}, \bibinfo{title}{Unraveling deformation sensing behavior of 3d printed continuous carbon fiber composites: Insights from fiber breakage and contact mechanisms}, \bibinfo{year}{2025}. \DOIprefix\doi{10.2139/ssrn.5396512}.
\bibitem[{Liu et~al.(2025)Liu, Miao, Deng, Hu, Zhang, Wang, Wang, Su, and Mai}]{LIU2025}
\bibinfo{author}{L.~Liu}, \bibinfo{author}{Y.~Miao}, \bibinfo{author}{Q.~Deng}, \bibinfo{author}{X.~Hu}, \bibinfo{author}{Y.~Zhang}, \bibinfo{author}{R.~Wang}, \bibinfo{author}{Y.~Wang}, \bibinfo{author}{M.~Su}, \bibinfo{author}{Y.-W. Mai},
\newblock \bibinfo{title}{Rate-dependent mechanical and self-monitoring behaviors of 3d printed continuous carbon fiber composites},
\newblock \bibinfo{journal}{Composites Science and Technology} \bibinfo{volume}{259} (\bibinfo{year}{2025}) \bibinfo{pages}{110914}. \URLprefix \url{https://www.sciencedirect.com/science/article/pii/S0266353824004846}. \DOIprefix\doi{https://doi.org/10.1016/j.compscitech.2024.110914}.
\bibitem[{Banks and Emami(0)}]{Banks_review}
\bibinfo{author}{J.~D. Banks}, \bibinfo{author}{A.~Emami},
\newblock \bibinfo{title}{Carbon-based piezoresistive polymer nanocomposites by extrusion additive manufacturing: Process, material design, and current progress},
\newblock \bibinfo{journal}{3D Printing and Additive Manufacturing} \bibinfo{volume}{0} (\bibinfo{year}{0}) \bibinfo{pages}{null}. \URLprefix \url{https://doi.org/10.1089/3dp.2022.0153}. \DOIprefix\doi{10.1089/3dp.2022.0153}.
\bibitem[{Schouten et~al.(2021)Schouten, Wolterink, Dijkshoorn, Kosmas, Stramigioli, and Krijnen}]{Schouten2020}
\bibinfo{author}{M.~Schouten}, \bibinfo{author}{G.~Wolterink}, \bibinfo{author}{A.~Dijkshoorn}, \bibinfo{author}{D.~Kosmas}, \bibinfo{author}{S.~Stramigioli}, \bibinfo{author}{G.~Krijnen},
\newblock \bibinfo{title}{A review of extrusion-based 3d printing for the fabrication of electro- and biomechanical sensors},
\newblock \bibinfo{journal}{IEEE Sensors Journal} \bibinfo{volume}{21} (\bibinfo{year}{2021}). \DOIprefix\doi{10.1109/JSEN.2020.3042436}.
\bibitem[{Zhou et~al.(2022)Zhou, Lian, Li, Yin, Ji, Li, Qi, and Huang}]{zhou2022}
\bibinfo{author}{Y.~Zhou}, \bibinfo{author}{H.~Lian}, \bibinfo{author}{Z.~Li}, \bibinfo{author}{L.~Yin}, \bibinfo{author}{Q.~Ji}, \bibinfo{author}{K.~Li}, \bibinfo{author}{F.~Qi}, \bibinfo{author}{Y.~Huang},
\newblock \bibinfo{title}{Crack engineering boosts the performance of flexible sensors},
\newblock \bibinfo{journal}{VIEW} \bibinfo{volume}{3} (\bibinfo{year}{2022}) \bibinfo{pages}{20220025}. \URLprefix \url{https://onlinelibrary.wiley.com/doi/abs/10.1002/VIW.20220025}. \DOIprefix\doi{https://doi.org/10.1002/VIW.20220025}.
\bibitem[{{Anisoprint SARL}(2022)}]{CCF_Datasheet}
\bibinfo{author}{{Anisoprint SARL}}, \bibinfo{title}{Anisoprint cfc petg material technical data sheet}, \bibinfo{year}{2022}. \URLprefix \url{https://anisoprint.com/wp-content/uploads/2022/03/CFC_PETG_datasheet_March_2022.pdf}.
\bibitem[{{Anisoprint}(2024)}]{Aura}
\bibinfo{author}{{Anisoprint}}, \bibinfo{title}{Aura 2 slicer}, \bibinfo{year}{2024}. \URLprefix \url{https://anisoprint.com/aura/}, \bibinfo{note}{accessed at 2024-03-12}.
\bibitem[{Heikkinen et~al.(2018)Heikkinen, Kauppinen, Liu, Asikainen, Spoljaric, Seppälä, Savin, and Pearce}]{PETG_chemical_stability}
\bibinfo{author}{I.~T. Heikkinen}, \bibinfo{author}{C.~Kauppinen}, \bibinfo{author}{Z.~Liu}, \bibinfo{author}{S.~M. Asikainen}, \bibinfo{author}{S.~Spoljaric}, \bibinfo{author}{J.~V. Seppälä}, \bibinfo{author}{H.~Savin}, \bibinfo{author}{J.~M. Pearce},
\newblock \bibinfo{title}{Chemical compatibility of fused filament fabrication-based 3-d printed components with solutions commonly used in semiconductor wet processing},
\newblock \bibinfo{journal}{Additive Manufacturing} \bibinfo{volume}{23} (\bibinfo{year}{2018}) \bibinfo{pages}{99--107}. \URLprefix \url{https://www.sciencedirect.com/science/article/pii/S2214860418303476}. \DOIprefix\doi{https://doi.org/10.1016/j.addma.2018.07.015}.
\bibitem[{{Protoplant Inc.}(2023)}]{Protopasta}
\bibinfo{author}{{Protoplant Inc.}}, \bibinfo{title}{Protopasta conductive {PLA}}, \bibinfo{year}{2023}. \URLprefix \url{https://www.proto-pasta.com/pages/conductive-pla}.
\bibitem[{Hibbeler(2017)}]{Hibbeler_MoM}
\bibinfo{author}{R.~Hibbeler}, \bibinfo{title}{Mechanics of Materials, SI Edition}, \bibinfo{publisher}{Pearson Education}, \bibinfo{year}{2017}. \bibinfo{note}{Chapter 6: Bending}.
\bibitem[{Lee and Liu(2019)}]{delta_t_ambient}
\bibinfo{author}{C.-Y. Lee}, \bibinfo{author}{C.-Y. Liu},
\newblock \bibinfo{title}{The influence of forced-air cooling on a 3d printed pla part manufactured by fused filament fabrication},
\newblock \bibinfo{journal}{Additive Manufacturing} \bibinfo{volume}{25} (\bibinfo{year}{2019}) \bibinfo{pages}{196--203}. \URLprefix \url{https://www.sciencedirect.com/science/article/pii/S2214860418306924}. \DOIprefix\doi{https://doi.org/10.1016/j.addma.2018.11.012}.
\bibitem[{Mirdehghan(2021)}]{fiber_strength}
\bibinfo{author}{S.~A. Mirdehghan},
\newblock \bibinfo{title}{1 - fibrous polymeric composites},
\newblock in: \bibinfo{editor}{M.~Latifi} (Ed.), \bibinfo{booktitle}{Engineered Polymeric Fibrous Materials}, The Textile Institute Book Series, \bibinfo{publisher}{Woodhead Publishing}, \bibinfo{year}{2021}, pp. \bibinfo{pages}{1--58}. \URLprefix \url{https://www.sciencedirect.com/science/article/pii/B9780128243817000123}. \DOIprefix\doi{https://doi.org/10.1016/B978-0-12-824381-7.00012-3}.
\bibitem[{Kulkarni and Ochoa(2006)}]{CTE_carbon_fiber}
\bibinfo{author}{R.~Kulkarni}, \bibinfo{author}{O.~Ochoa},
\newblock \bibinfo{title}{Transverse and longitudinal cte measurements of carbon fibers and their impact on interfacial residual stresses in composites},
\newblock \bibinfo{journal}{Journal of Composite Materials} \bibinfo{volume}{40} (\bibinfo{year}{2006}) \bibinfo{pages}{733--754}. \URLprefix \url{https://doi.org/10.1177/0021998305055545}. \DOIprefix\doi{10.1177/0021998305055545}. \href{http://arxiv.org/abs/https://doi.org/10.1177/0021998305055545}{{\tt arXiv:https://doi.org/10.1177/0021998305055545}}.
\bibitem[{Plastics(2013)}]{PETG_CTE}
\bibinfo{author}{B.~Plastics}, \bibinfo{title}{Petg technical datasheet}, \bibinfo{year}{2013}. \URLprefix \url{https://www.theplasticpeople.co.uk/ThePlasticPeople/media/The-Plastic-People/Products/Cut%20To%20Size/Data%20Sheets/datasheet-petg-2014.pdf?ext=.pdf}.
\bibitem[{Polymers(2023)}]{PETG_TDS}
\bibinfo{author}{P.~Polymers}, \bibinfo{title}{Prusament petg technical datasheet}, \bibinfo{year}{2023}. \URLprefix \url{https://prusament.com/wp-content/uploads/2023/07/PETG_V0_ENG.pdf}.
\bibitem[{Polymers(2022)}]{PLA_TDS}
\bibinfo{author}{P.~Polymers}, \bibinfo{title}{Prusament pla technical datasheet}, \bibinfo{year}{2022}. \URLprefix \url{https://prusament.com/wp-content/uploads/2022/10/PLA_Prusament_TDS_2021_10_EN.pdf}.
\bibitem[{Iizuka and Todoroki(2023)}]{Iizuka}
\bibinfo{author}{K.~Iizuka}, \bibinfo{author}{A.~Todoroki},
\newblock \bibinfo{title}{Nonlinear behavior mechanism of change in electrical resistance on 3dprinted carbon fiber / pa6 composites during cyclic tests},
\newblock \bibinfo{journal}{Advanced Composite Materials} \bibinfo{volume}{32} (\bibinfo{year}{2023}) \bibinfo{pages}{1--20}. \URLprefix \url{https://doi.org/10.1080/09243046.2022.2055514}. \DOIprefix\doi{10.1080/09243046.2022.2055514}.
\bibitem[{Luan et~al.(2018)Luan, Yao, Liu, Lan, and Fu}]{fracture_detection}
\bibinfo{author}{C.~Luan}, \bibinfo{author}{X.~Yao}, \bibinfo{author}{C.~Liu}, \bibinfo{author}{L.~Lan}, \bibinfo{author}{J.~Fu},
\newblock \bibinfo{title}{Self-monitoring continuous carbon fiber reinforced thermoplastic based on dual-material three-dimensional printing integration process},
\newblock \bibinfo{journal}{Carbon} \bibinfo{volume}{140} (\bibinfo{year}{2018}) \bibinfo{pages}{100--111}. \URLprefix \url{https://www.sciencedirect.com/science/article/pii/S0008622318307498}. \DOIprefix\doi{https://doi.org/10.1016/j.carbon.2018.08.019}.

\end{thebibliography}

\end{document}